\documentclass[twoside,final]{pro12}

\usepackage{amsmath}
\usepackage{tabularx,multirow,hhline}
\usepackage{graphicx}

\usepackage{lineno}

\head{L. Lamy et al.} {SKR before the Cassini Grand Finale}	

\begin{document}

\title{THE SATURNIAN KILOMETRIC RADIATION BEFORE THE CASSINI GRAND FINALE}	
\author{L. Lamy\adress{\textsl LESIA, Observatoire de Paris, CNRS, PSL, UPMC, UPD, Meudon, France}}

\maketitle

\begin{abstract}

The Saturnian Kilometric Radiation (SKR) is radiated from the auroral regions surrounding the kronian magnetic poles, above the ionosphere up to a few planetary radii. It directly compares to the auroral radio emissions emanating from other planetary magnetospheres such as the Earth and the giant planets. Our knowledge on SKR relied on remote observations of Voyager (flybys in 1980 and 1981) and Ulysses (distant observations in the 1990s) until Cassini started to orbit Saturn in 2004. Since then, it has been routinely observed from a large set of remote locations, but also in situ for the first time at a planet other than Earth. This article reviews the state of the art of SKR average remote properties, the first insights brought by in situ passes within its source region, together with some remaining questions before the Cassini Grand Finale and its close-in polar orbits.


\end{abstract}

\section{Introduction}

Following Jupiter and the Earth, Saturn was the third planet found to host non-thermal auroral radio emissions at low frequencies, typically ranging from a few kHz to 1~MHz. The Saturnian Kilometric Radiation (SKR) was first unambiguously detected by the Voyager mission [Kaiser et al., 1980] with the Planetary Radio Astronomy (PRA) instrument connected to two 10~m long monopoles. The Voyager 1 \& 2 spacecraft measured SKR from distances to Saturn as far as 4~AU (astronomical units) before successively flying by the planet in November 1980 and August 1981. SKR was then occasionally detected by the Ulysses spacecraft at (very) large distances from the planet, $\ge5$~AU, between 1993 and 2002 with the Unified Radio and Plasma Wave (URAP) experiment using a 72~m tip-to-tip dipole and a 7.5~m long monopole [Galopeau et al., 2000]. With the Radio and Plasma Wave Science (RPWS) experiment connected to three 10~m long monopoles, the Cassini mission started to observe SKR in 2003, from several AU, before the orbit insertion in mid-2004 [Kurth et al., 2005; Gurnett et al., 2005]. Since then, SKR has been observed quasi-continuously from a wide variety of locations around the planet.

The SKR remote properties have been first established from Voyager/PRA observations [Kaiser et al., 1984 and references therein] before to be statistically investigated from Cassini/RPWS ones [Lamy et al., 2008a]. These characteristics have been summarized in recent successive reviews [Kurth et al., 2009; Badman et al., 2015, Stallard et al., in press] and are only briefly reminded below. The kronian kilometric radiation is the most powerful component of Saturn's radio spectrum and strikingly compares to the terrestrial Auroral Kilometric Radiation (AKR) in terms of spectral range and intensity. Radio waves are radiated at $f\sim f_{ce}$ (the local electron gyrofrequency) along high latitudes magnetic field lines colocated with the auroral oval. SKR propagates mainly and marginally on the extraordinary (R-X) and ordinary (L-O) free-space modes, respectively, at large angles from the local magnetic field direction. Kilometric waves are fully elliptically polarized, with right-handed (RH) and left-handed (LH) polarizations corresponding to R-X mode (L-O mode, respectively) emission radiated from the northern and southern (southern and northern, respectively) hemispheres. Overall, these properties were early shown to be consistent with those expected from the Cyclotron Maser Instability (CMI), the mechanism driving AKR at Earth and suspected to similarly apply to all planetary auroral radio emissions [Wu and Lee, 1979; Zarka, 1998; Treumann, 2006]. This inference could then be validated in situ in 2008 when the Cassini spacecraft unexpectedly crossed the SKR source region, for the first time at a planet other than Earth [Lamy et al., 2010; Mutel et al., 2010; Menietti et al., 2011]. The Cassini mission will culminate in 2017 with the Grand Finale, a series of polar orbits which will end with the final dive of the spacecraft into Saturn's atmosphere on 15 September 2017. These polar passes are a unique opportunity to repeatedly sample in situ the auroral kilometric sources of both hemispheres over a wide range of frequencies, distances and local times (LT) and assess how the CMI operates in Saturn's auroral regions and how the wave properties evolve from its source region to the observer.

This article reviews our current knowledge of SKR main remote (section \ref{remote}) and local (section \ref{local}) properties together with some remaining questions before the Cassini end of mission (section \ref{perspectives}). 

\section{Remote characteristics}
\label{remote}

\subsection{Spectrum}
\label{spectrum}

\begin{figure}[ht]
\centering
\includegraphics[width=1\textwidth]{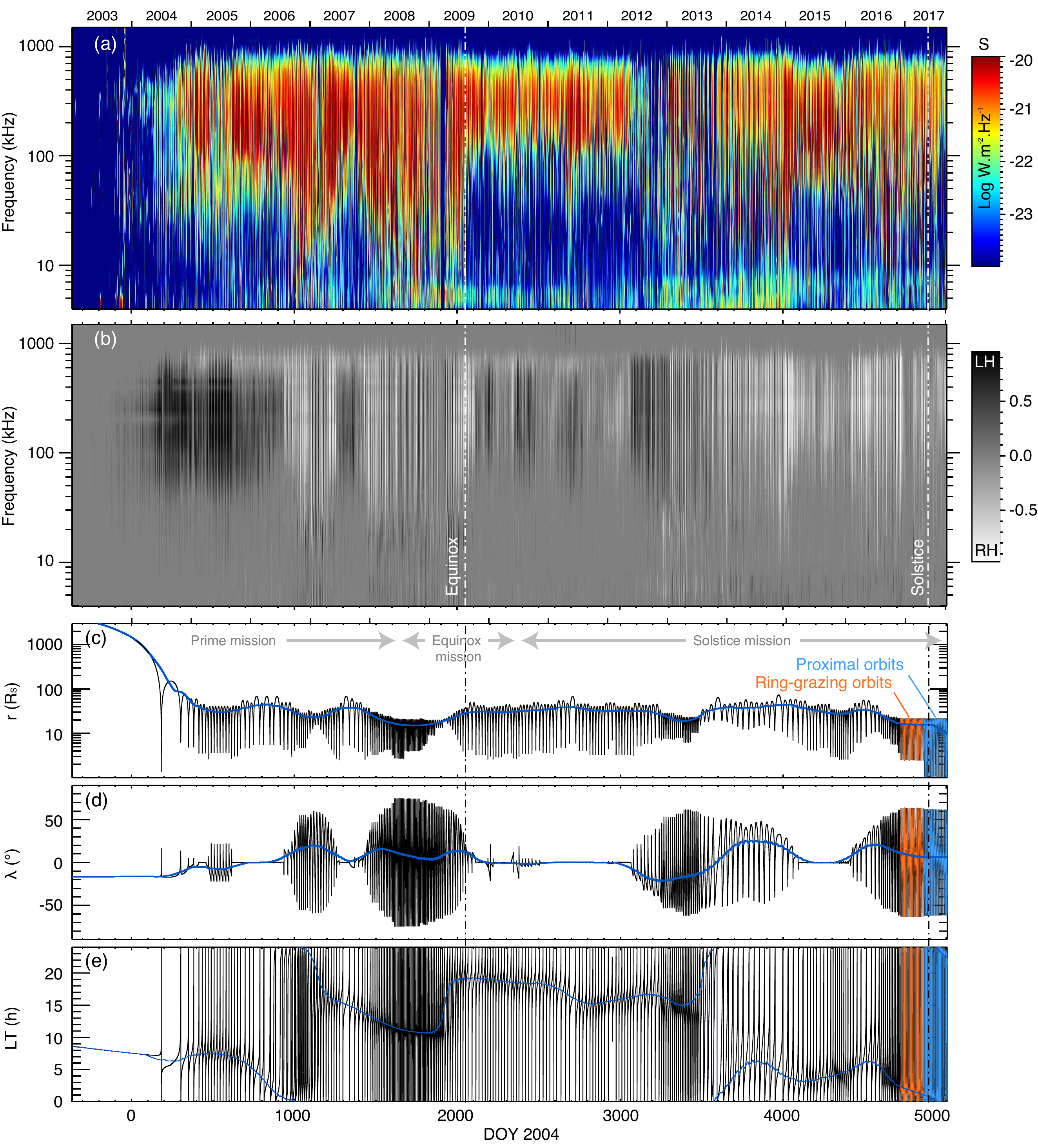}
\caption{Cassini/RPWS observations of SKR (a) spectral flux density normalized at 1~AU and (b) degree of circular polarization V ranging from 2003-001 to 2017-258, as processed in [Lamy et al., 2008a]. Cassini's planetocentric (c) distance to the planet, (d) latitude and (e) local time. Blue lines indicate sliding averages integrated over 200 days.}
\label{fig1}
\end{figure}

The SKR R-X mode spectrum can extend from 1200~kHz down to $-$ and below $-$ 10~kHz, which transposes in distances above the ionosphere up to $-$ and beyond $-$ 5 planetary radii (1~R$_S$~=~60268~km). It peaks over 100-400~kHz with a flux density occasionally reaching $10^{-19}$~Wm$^{-2}$Hz$^{-1}$ at 1~AU, for the 1\% occurrence level. L-O mode SKR has also been identified but with intensities generally lower by two orders of magnitude. Therefore, unless otherwise mentioned, SKR hereafter refers to R-X mode SKR. The northern SKR spectrum reaches slightly higher frequencies than the southern one (typically $\sim$100~kHz), owing to the 0.04~R$_S$ northward magnetic field offset. The statistical investigation of SKR spectrum as a function of Cassini's location and hemisphere of origin illustrates a strong variability. Overall, the radiation is most intense when observed from the dawn sector and from moderate latitudes for a given hemisphere. Waves from both hemispheres can nonetheless be observed simultaneously from near-equator latitudes up to $\pm20^\circ$, where they can still be discriminated by polarization [Kaiser et al., 1981; Lamy et al., 2008a]. SKR northern and southern spectra also vary as a function of time at different time scales (see section \ref{dynamics}), with the most intense radiation coming from the summer hemisphere [Kimura et al., 2013]. Figure \ref{fig1} provides an illustration of RPWS observations of SKR over the course of the whole Cassini mission. The total flux density (panel a) displays wideband variations mainly related to orbital effects (panels c to e). Considering Cassini near-equatorial passes, the predominant hemisphere changed roughly 2 years past the equinox of 2009 : southern SKR (LH polarized on panel b) was predominant in late 2005-early 2006, in late 2007 and in 2010-mid 2011, while the northern SKR (RH polarized) was predominant in late 2011 and in 2015. 

Galopeau et al. [1989] attempted to theoretically model the SKR spectrum. In the absence of any in situ detailed measurements of the auroral region at that time, they assumed the CMI as a source mechanism operating in an inhomogeneous medium ($i.e.$ the spatial variation of plasma parameters is not negligible with respect to the wavelength), they hypothesized non-linear wave saturation by trapping and used macroscopic plasma parameters in agreement with the observations (magnetospheric magnetic field, plasma density, characteristic energy of unstable electrons) to derive a maximum wave electric field. The calculated theoretical spectrum was in good agreement with the observations. Precisely, the modelled spectral intensities were larger than the most intense observed ones by one order of magnitude, suggesting that SKR is only marginally saturated by nonlinear processes. Among the first set of studies which later investigated the SKR source region (see section \ref{local}), the auroral medium was then found to be rather homogeneous [Lamy et al., 2011].


\subsection{Locus of radio sources}

Using independent methods to analyze Voyager/PRA measurements, the locus of kilometric sources was first indirectly constrained from visibility conditions to an active region at 08:00-13:00 Local Time (LT) at latitudes $\ge70^\circ$ in both hemispheres [Kaiser and Desch, 1982; Lecacheux and Genova, 1983; Galopeau et al., 1995]. Since SKR activity was found to be correlated with that of atmospheric aurorae and because AKR sources lie along field lines mapping to bright spots of the terrestrial auroral oval [Huff et al., 1988], the high resolution images of Saturn's northern aurorae showing a circumpolar auroral oval [e.g. Trauger et al., 1998] early raised the question of SKR sources more widely distributed in longitude. 

Cassini/RPWS measurements of SKR between 100 and 300~kHz acquired close to midnight on day 2004-183 enabled [Farrell et al., 2005] to infer a nightside SKR source region. Cecconi et al. [2009] then led a direction-finding $-$ or goniopolarimetric $-$ analysis of RPWS data acquired around the perikrone of day 2006-268. This technique directly provides the wave polarization (Stokes parameters S,Q,U and V) and Poynting vector \textbf{k} at each time-frequency measurement and, assuming a magnetic field model, emission at $f_{ce}$ and straight line propagation, the 3D locus of the radio source together with its associated magnetic field line (and footprint) and the aperture angle $\theta$ = (\textbf{k},\textbf{B}) at the source, where \textbf{B} is the local magnetic field vector. Cecconi et al. [2009] located the magnetic footprint of SKR sources at all frequencies, observed in both R-X and L-O modes, over a region covering $70^\circ$ to $80^\circ$ latitude in both hemispheres and  04:00-16:00~LT. Lamy et al. [2009] extended this study by building radio maps of SKR data at all frequencies observed between mid-2004 and early 2008 and showed that SKR sources actually lie along a circumpolar oval covering all local times and colocated with the UV main auroral oval in latitude and longitude, both statistically and instantaneously. The intensity of SKR (and UV) emission varies with LT and peaks at 08:00~LT, with a full width at half maximum of $\sim$ 08:00~LT. 

The spatial association between SKR sources and UV aurorae led to the conclusion that a common electron population should feed both auroral processes. Nevertheless, the UV (but also IR and visible) aurorae have since then witnessed a large variety of components beyond the so-called main oval, including polar emissions, a secondary equatorward oval [Badman et al., this issue] etc., which might have a radio counterpart as well, as long as they involve mildly relativistic electrons able to drive CMI [Treumann, 2006]. A statistical analysis of the energy of precipitating electrons driving Saturn's aurorae yield a typical 1-17keV range [Gustin et al., 2017] which support such an association.

\subsection{Beaming pattern}
\label{beaming}

Kilometric waves are radiated along a thin hollow cone, which in turn drives strong visibility effects that produce (very) structured emission in the time-frequency plane, with bursts of various durations and shapes, and render the interpretation of radio observations difficult. As mentioned in section \ref{spectrum}, SKR is mainly beamed toward its hemisphere of origin but can be observed from the other one up to 20$^\circ$ latitude [Lamy et al., 2008a; Kimura et al., 2013]. Measuring precisely the aperture (or beaming) angle $\theta$ at the source $-$ hereafter defined as (\textbf{k},\textbf{B}) in the north and (\textbf{k},\textbf{-B}) in the south for comparison purposes $-$ is of particular interest as it provides a direct remote constraint of the local wave amplification. $\theta$ has historically been determined through two distinct approaches.

\begin{figure}[ht]
\centering
\includegraphics[width=0.5\textwidth]{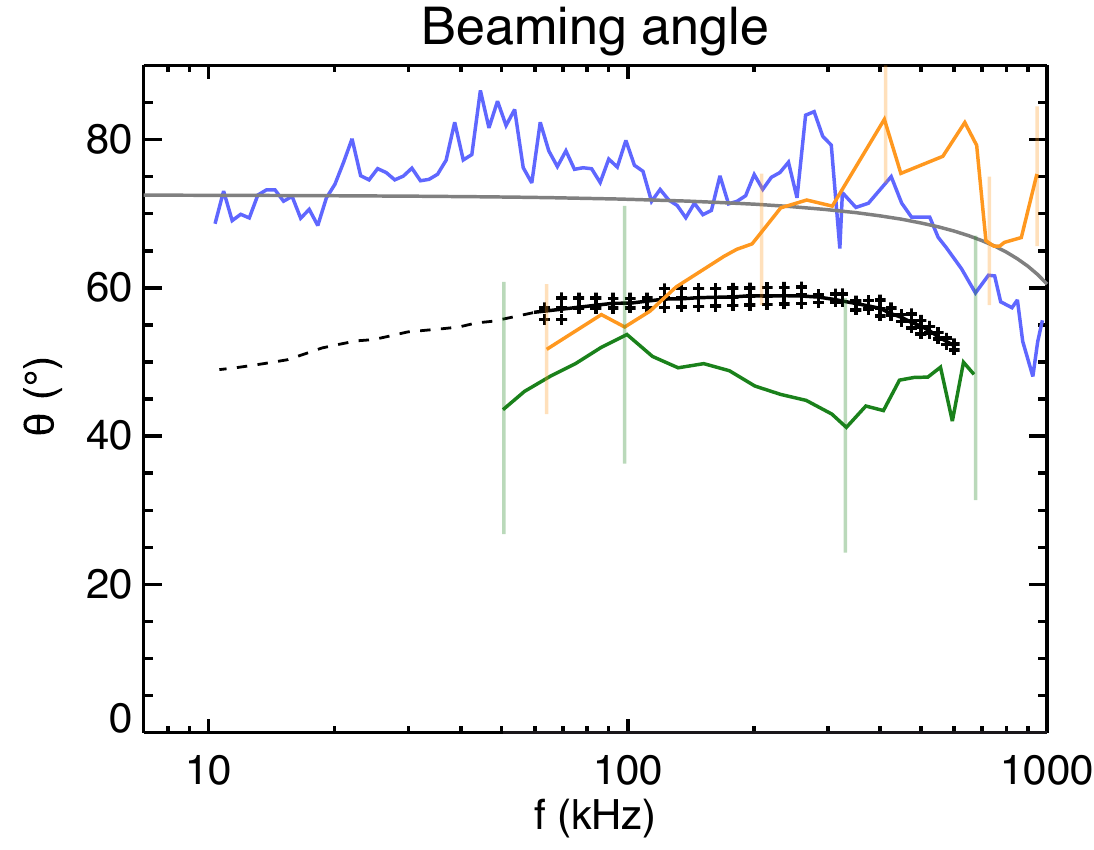}
\caption{SKR beaming angles as derived from Cassini/RPWS observations either indirectly by [Lamy et al., 2008b] (solid gray line) or directly by [Cecconi et al., 2009] (solid green and orange lines, with vertical error bars), [Lamy et al., 2011] (solid blue line) and [Lamy et al., 2013] (solid and dashed black lines). The orange solid line refers to northern emissions and the other lines to southern ones.}
\label{fig2}
\end{figure}

It can first be indirectly inferred from modelling studies. Thieman et al. [1981] for instance used an ad hoc model of hollow conical sheet, using $\theta(f)=75-85^\circ$ for radio sources along field lines of apex $\le5$~radii, to reproduce the curved kilometric arcs regularly observed in Voyager/PRA dynamic spectra [Boischot et al., 1981]. Lamy et al. [2008b, 2008c] quantitatively simulated SKR visibility effects frequently observed in Cassini/RPWS dynamic spectra, such as double RH/LH polarized arcs and signal extinctions close to the planet or at latitudes beyond 60$^\circ$ in both hemispheres, assuming a CMI-driven emission angle, radio sources colocated with the main auroral oval (field line apex around 12-15~radii) and straight line propagation. They found that the observations were best reproduced with an oblique emission ($\theta$ decreasing with frequency, solid gray line in Figure \ref{fig2}) driven by a loss cone electron distribution using 20~keV electrons. Using a self-consistent beaming angle (solid and dashed black lines in Figure \ref{fig2}) determined from the fit of a well-defined SKR southern arc conjugate with a UV sub-corotating auroral spot observed simultaneously, Lamy et al. [2013] then simulated the southern SKR rotational modulation with an active region of the main auroral oval, 90$^\circ$ wide in longitude, rotating at the southern SKR period. 

The direct measurement of $\theta$ was made possible by the goniopolarimetric analysis of Cassini/RPWS data, whose principle has been described above. Investigating R-X mode SKR, Cecconi et al. [2009] measured $\theta$ between 50$^\circ$ and 80$^\circ$ in the northern hemisphere and between 45$^\circ$ and 55$^\circ$ in the southern hemisphere (respectively displayed by solid orange and green lines in Figure \ref{fig2}) from 50-60 to 700~kHz. Lamy et al. [2011] studied R-X mode emission observed close to the southern SKR source region and measured $\theta$ varying from 80$^\circ$ to 50$^\circ$ from 10 to 1000~kHz (solid blue line in Figure \ref{fig2}). As detailed in section \ref{local}, these authors obtained the first measurements of $\theta$ within the source region itself, found to be consistent with a perpendicular emission.

These different estimates of SKR beaming angles are summarized and compared in Figure \ref{fig2}. They illustrate that, when observed remotely, the apparent emission angle is oblique and does not only vary with frequency but also with time and/or observer's location. Understanding how the (assumed) initial perpendicular emission angle at the source becomes oblique with propagation is necessary to correctly assess the information brought by $\theta$ onto the source region. Studies of the terrestrial AKR or Jupiter DAM emission cones [e.g. Mutel et al., 2004 ; Galopeau et al., 2016] show that refraction effects at/near the source, neglected in the above mentioned studies, are likely to drive an azimuthal asymmetry of $\theta$ around the local magnetic field vector \textbf{B}.

\subsection{Polarization}
\label{polar}

Warwick et al. [1981, 1982] and Ortega-Molina \& Lecacheux [1990] reported that SKR observed along the ecliptic trajectories of Voyager 1 \& 2 was nearly purely circularly polarized with predominant (marginal) RH (LH) polarized waves originating from the northern (southern) hemisphere, as expected for R-X mode emission. The measurement of fully circularly polarized waves (Stokes parameters V$\sim$1 and U,Q$\sim$0) in both L-O and R-X modes was later confirmed with near-equatorial and statistical 2-antenna Cassini/RPWS data acquired from 2004 to early 2007 [Cecconi et al., 2006; Lamy et al., 2008a].

However, the subsequent detailed analysis by [Fischer et al., 2009] of high latitude 3-antenna measurements ranging from late 2006 to early 2007 revealed that, if SKR waves are indeed fully polarized (namely T~$=\sqrt{U^2+Q^2+V^2}=1$), the polarization is rather circular below $30^\circ$ latitude and elliptical beyond with an increasing amount of linear polarization (L~$=\sqrt{U^2+Q^2}\ne0$) with increasing latitude. The authors discussed various possible theoretical interpretations involving mode coupling able to change or freeze the wave polarization along the raypath from the source or a limiting polarization zone.

Anticipating the results presented in section \ref{local}, 3-antenna polarization measurements within the SKR source region were later analyzed by [Lamy et al., 2011]. The authors found that the polarization of both R-X and L-O mode emission is almost fully linear at the source (L$\sim$1, V$\sim$0), as expected for a quasi-perpendicular emission, and evolves toward circular polarization along the propagation away from the source (V$\sim$1 being reached after a $\sim$2~radii long path). This behaviour was shown to be fully consistent with that predicted by the magnetoionic theory in a cold homogenous plasma such as that observed in the auroral region. Precisely, when the medium is varying slowly enough along the raypath, the characteristic magnetoionic modes are weakly coupled and the polarization of each is defined by (and slowly changes with) the local plasma parameters.

A final area of interest concerning SKR polarization is that only a few rare examples of Faraday rotations have been observed so far (A. Lecacheux, personal communication). If weak mode coupling conditions are indeed fullfilled in most of the magnetosphere, the wave polarization is expected to continuously evolve in its characteristic mode while propagating away from the source. Conversely, observing Faraday rotation would imply that a wave initially propagating in a pure mode crossed at some point a strongly inhomogeneous plasma layer, during which the polarization was frozen by strong mode coupling conditions, and after which the wave polarization ellipse could then decompose into a mixture of R-X or L-O modes able to propagate at their own velocity to produce the rotation of the ellipse polarization plane known as Faraday rotation.

\subsection{Dynamics}
\label{dynamics}

The kronian auroral radio emissions display a rich type of dynamics, at short and long time scales, besides the SKR rotational modulation period of $\sim10.7$~h specifically dealt with in the next section.

At time scales lower than a few hours, SKR displays numerous arc-shaped structures [Boischot et al., 1981], which are understood as the rotational motion of active flux tubes carrying radio sources distributed over a wideband spectral range. Their shape is vertex early (late), which means that the center of curvature is located right (left) to the arc, whenever the active flux tube enters (exits) the field-of view of a near-equatorial observer. The arc curvature thus provides a direct diagnostic of the spatial locus of sources with respect to the observer. Interestingly, the accurate modelling of several of these arcs showed a sub-corotational motion of flux tubes carrying radio sources rather than a rigid corotation, as observed for isolated auroral spots [Lamy et al., 2008b, Lamy et al., 2013] or equivalently for Jupiter decametric emissions [e.g. Hess et al., 2014]. At time scales ranging from minutes to second, [Kurth et al., 2005] identified a variety of SKR fine structures, with bandwidths down to 200~Hz and drifting in frequency either with positive or negative slopes at rates of a few kHz per second (hence speeds of a few km per second). Such fine structures are reminiscent of those observed for AKR, but their origin (for example solitary ion structures, electromagnetic ion cyclotron waves) remains an open question [Kopf and Gurnett, 2011].

At time scales larger than a few rotations, SKR displays global intensifications, especially at low frequencies which can last for a few hours to a few days. The long-lasting intensifications have been early related to sudden rises of the solar wind dynamic pressure and associated with dawnside auroral storms [Desch, 1982 ; Desch and Rucker, 1983a, 1985 ; Rucker et al., 2008; Kurth et al., 2016]. The recurrent passes of Saturn's magnetosphere within Jupiter's distant magnetotail which coincided with abrupt dropouts of SKR intensity strikingly illustrated the crucial role of solar wind in driving auroral kilometric emission [Desch, 1983a]. SKR low frequency extensions have also been related to plasmoid ejections in the magnetotail related to substorm-like activity [Jackman et al., 2009]. A case study of one short-lasting low frequency SKR intensification observed during quiet solar wind conditions suggested that it was driven internally by the planetary rotation [Lamy et al., 2013]. Interestingly, global magnetospheric compressions which trigger long-lasting SKR enhancements are also accompanied by tailward plasmoid ejection [Jackman et al., 2010]. More detailed reviews of the SKR response to magnetospheric compressions and tail reconnection can be found in [Jackman et al., 2011; Jackman et al. this issue].

\subsection{Rotational modulation}

The ubiquitous rotational modulation of magnetospheric observables (auroral emissions, magnetic field, plasma parameters) at two different, variable, hemispheric periods has been widely studied through an extensive literature [see e.g. the review of Carbary et al., in press]. Although these hemispheric periodicities are interpreted as the result of two global rotating systems of magnetospheric field-aligned currents [e.g. Andrews et al., 2010 ; Southwood and Cowley, 2014] likely produced by atmospheric vortical flows [Jia et al., 2012; Hunt et al., 2014], the ultimate physical origin of such current systems remains unknown and, as a result, an active research topic of the end of the Cassini mission. 

Different methods have been used to determine and analyze SKR hemispheric periods and derive successive longitude/phase systems with Voyager, Ulysses and Cassini data. The reader interested in a brief summary of the results obtained until 2010 is referred to [Lamy, 2011; Gurnett et al., 2011 and references therein].

Since then, the post-equinox evolution of SKR periods, similarly observed in magnetic oscillations, auroral hiss and narrow band emissions, has been investigated through several studies. Over the 2010-2012 interval, SKR periods were found to remain very close to each other with locked phases [Provan et al., 2014; Fischer et al., 2015]. [Fischer et al., 2014] noticed occasional unusual abrupt changes of periods and phaseshift in 2011 which they tentatively attributed to the Saturn's Great White Spot activity, or alternately to varying solar wind conditions by Provan et al. [2015]. The identification of these unusual events was nonetheless questioned by [Cowley and Provan, 2015] based on a separate analysis of SKR and magnetic periods which had yielded different results [Provan et al., 2014]. Both SKR (and other magnetic or radio) periods eventually merged between mid-2013 and mid-2014 before crossing and diverging from each other after mid-2014 [Provan et al., 2016; Ye et al., 2016; Ye et al., this issue]. 

\begin{figure}[ht]
\centering
\includegraphics[width=1\textwidth]{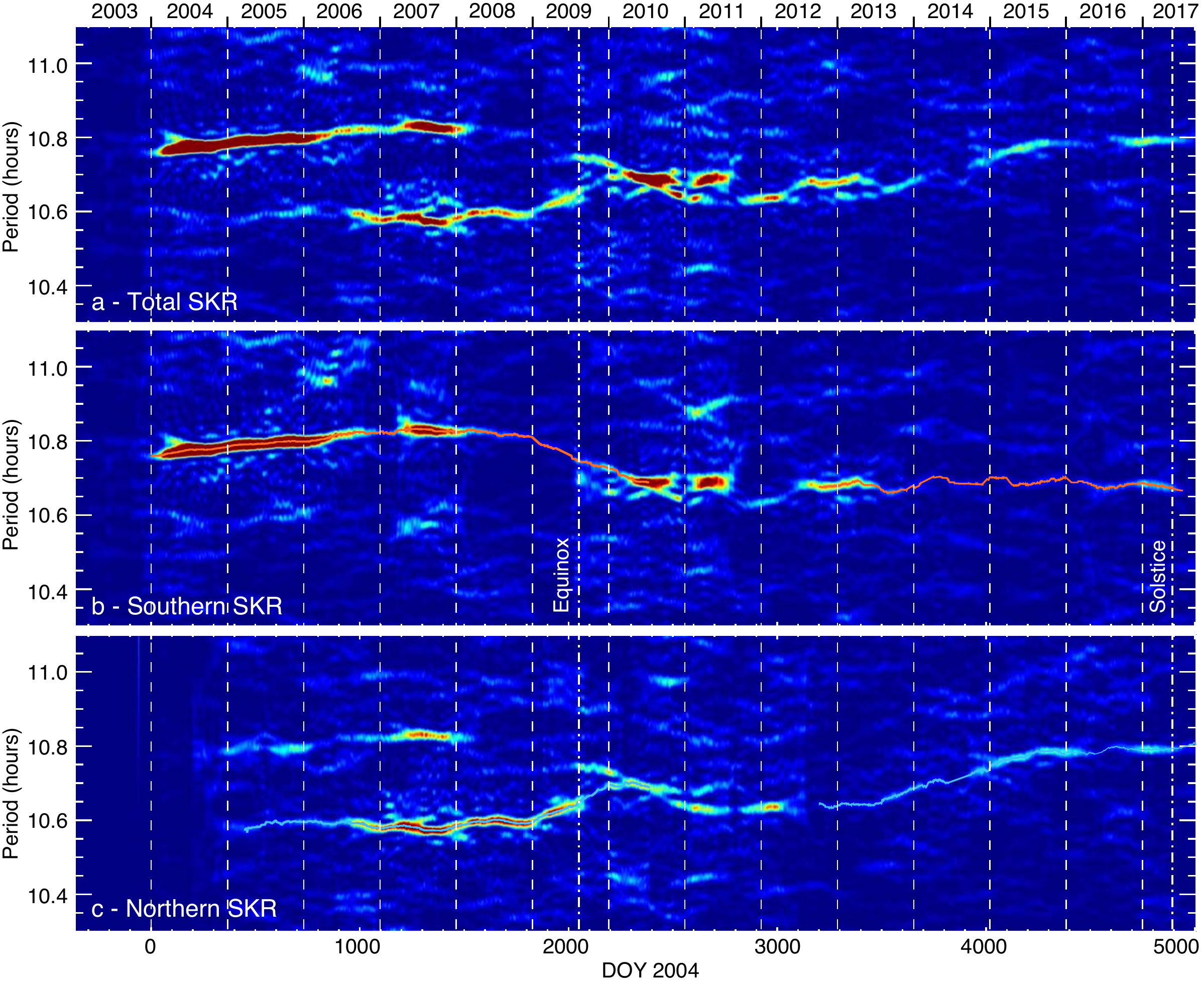}
\caption{Lomb-Scargle periodograms of (a) total, (b) southern and (c) northern SKR flux density integrated over 100-500~kHz as a function of period and time from 2003-001 to 2017-258, extending Fig. 1 of [Lamy, 2011]. Southern and northern SKR were originally identified as LH and RH (X-mode) emission, excluding scarce intervals displaying ambiguous polarization measurements. The latter intervals were noticed to be more frequent at close distances and high latitudes which prevailed in the past two years. To improve the signal-to-noise ratio of panels (b) and (c), southern and northern SKR were therefore identified as the total SKR signal observed beyond $10^\circ$ latitude after days 2016-001 and 2016-180, respectively. Solid red and blue lines plot the southern and northern SKR periods derived by [Lamy, 2011] from day 2004-001 to 2010-193, [Provan et al., 2016] from day 2012-278 to 2015-365 and tracked here from day 2016-001 to 2017-258. The latter will serve to build southern and northern SKR phases covering most of the Cassini mission.}
\label{fig3}
\end{figure}

Figure \ref{fig3} shows a summary of SKR periods as observed by RPWS over the whole Cassini mission. The three panels display Lomb periodograms of total, southern and northern SKR ranging from 2003 to 2017, extending Figure 1 of [Lamy, 2011]. Solid red and blue lines show the southern and northern SKR periods derived by [Lamy, 2011] (2004 to mid-2010), [Provan et al., 2016] (late 2012 to 2015) and tracked here up to day late 2017. Over 2016-2017, the northern period remained stable around 10.8~h, similarly to the southern period in 2007. Despite a lower signal-to-noise ratio, the southern period also appeared to remain stable close to 10.7~h from 2012 to 2016, and thus slower than the northern period in 2007 by 0.1~h. Interestingly, the southern SKR periodogram displays a secondary peak at the northern period after late 2016, a situation similar to the converse presence of the southern period in northern SKR in 2007.

Several possible causes have been proposed to account for the observed long-term period evolution, such as seasonal variations of the ionospheric conductivity [Gurnett et al., 2009] or a solar wind control of polar caps recently proposed as a possible driver of different hemispheric periods [Southwood, personal communication]. However, it is worth to note that the variation of SKR periods from the equinox of 1980 up to recent years [Gurnett et al., 2010] shows two episodes of non-monotonous variations : during the post-equinox interval of 2010-2012, when both periods remained closed to each other, and in 2002-2004, when the southern period experienced a marked decrease lasting for several years. Additionally, the southern SKR period reached an extremum in 2007, six years after the solstice and only two years before equinox, while the northern period was rather stable from 2005 to 2008, such as the southern period since 2012. These non-monotonous variations cannot be simply explained by the smoothed evolution of solar illumination or solar wind activity during these intervals.

\section{In situ properties}
\label{local}

In 2008, Cassini unexpectedly sampled the SKR source region on day 291 and possibly on day 73 as well. The analysis of radio (RPWS) and plasma (MAG, CAPS, INCA instruments) in situ measurements taken at these occasions provided the first clues on the generation of auroral radio emissions at a planet other than Earth.
 
 \subsection{Event 2008-291}

On day 2008-291, the spacecraft crossed the southern source region around 10~kHz ($\sim$5~R$_S$) at an unusual local time of 01:00 LT [Lamy et al., 2010; Kurth et al., 2011]. This atypical locus for kilometric sources matched an episode of global SKR intensification triggered by a solar wind-driven magnetospheric compression. Bunce et al. [2010] showed the presence of unusually intense field-aligned current signatures in the midnight and dawn sectors together with a polar cap contracted toward the pole associated with this event, resulting from a major open flux closure event in Saturn's tail. These current signatures were consistent with the spiral-shaped polar projection of SKR sources, at highest latitudes close to midnight and reminiscent of auroral storms [Lamy et al., 2010; Bunce, 2012].

The combined detailed analysis of radio and plasma measurements revealed both similarities and differences between the terrestrial and kronian cases. Three successive layers of SKR R-X mode sources (with nearby L-O mode sources) were identified along the spacecraft southern trajectory by RPWS measurements of the SKR low frequency cutoff $f_{cut}$ strictly below $f_{ce}$, with $(f-f_{ce})/f_{ce}$ as low as $-2$\% and source sizes of 900 to 1800~km. The auroral region displayed a tenuous plasma dominated by hot electrons and plasma frequencies $f_{pe}\le0.1f_{ce}$ as at Earth, but no terrestrial-like plasma cavities were observed within radio sources. This difference, together with the presence of SKR sources at larger distances than AKR sources at Earth, was attributed to the rapid rotation of Saturn's magnetosphere, which makes the $f_{pe}/f_{ce}\le 0.1$ condition necessary to CMI generally fullfilled at high latitudes. SKR sources were also found to match a layer of upward current, consistent with down-going electrons. Particle measurements revealed three populations of cold, warm and hot electrons. 

The CMI theoretical emission frequency $f_{CMI}$ was found to be consistent with $f_{cut}$ only if the emission is emitted quasi-perpendicularly and is driven by the hot electrons, observed to propagate downward with 6 to 9~keV energies (and thus implying an acceleration region located farther upward than radio sources along the flux tube) and shell-like distribution functions [Lamy et al., 2010; Kurth et al., 2011; Schippers et al., 2011]. By fitting the measured electron phase space density with a Dory-Guest-Harris (DGH) function and assuming linear source sizes of $\sim$1000~km, Mutel et al. [2010] computed CMI large enough growth rates to account for the observed SKR flux densities and perpendicular emission. Lamy et al. [2011] derived typical electric field amplitudes of 0.85 to 2.6~mV.m$^{-1}$ (1.7~mV.m$^{-1}$ in average) and electron-wave energy conversion efficiencies of 0.2 to 2\% (1\% in average).

Finally, as already introduced in sections \ref{beaming} and \ref{polar}, the goniopolarimetric analysis of 3-antenna RPWS measurements showed a strong elliptical polarization which gradually circularizes along the wave propagation, consistent with the magnetoionic theory and weak mode coupling conditions. RPWS direct measurements of $\theta$ also reveal quasi-perpendicular emission at the source, in agreement with estimates of $\theta$ through two independent methods : (i) the above mentioned agreement of $f_{cut}$ with $f_{CMI}$ assuming the observed 6-9~keV electrons as the CMI source of free energy and (ii) the direct measurement of the axial ratio of the polarization ellipse T~$\sim\cos\theta$ [Melrose and Dulk, 1991], with T~$\sim0.2$ implying $\theta\ge80^\circ$.

 \subsection{Event 2008-073}

On day 2008-073, Cassini went within or close to the radio source region around 4.5~kHz ($\sim$6~R$_S$) at a more usual dawn region of 08:00 LT, close the peak of SKR intensity in LT [Menietti et al., 2011a, 2011b]. R-X and LO mode emissions were observed with flux densities similar to those observed on day 2008-291. The measured $f_{pe}/f_{ce}$ ratio was as low as 0.05. These authors did not determine the emission's low frequency cutoff but directly investigated electron measurements to assess their compatibility with CMI-driven emission. They identified shell-like electron distribution around 7~keV, which they fitted with a DGH function to obtained high enough growth rates to account for the observed R-X mode SKR flux densities. They also obtained L-O mode flux densities much lower than observed and therefore identified the L-O mode feature as narrowband emission rather than SKR. 

\section{Conclusion and Perspectives}
\label{perspectives}

The Cassini mission will culminate with the Grand Finale in 2017 along two series of polar orbits : 20 ring-grazing orbits from November 2016 to April 2017 (Cassini revolutions 251 to 270 displayed by orange lines in Figure \ref{fig1}c-e, with perikrones just outside the F ring) followed by 22 proximal orbits from April to September 2017 (revolutions 271 to 293 displayed by blue lines in Figure \ref{fig1}c-e, with perikrones inside the rings) which will end with the final dive of the spacecraft into Saturn's atmosphere on 15 September 2017. These polar orbits will provide a unique opportunity to repeatedly sample in situ \textbf{auroral field lines} and SKR sources over a wide range of frequencies, distances and local times and therefore assess the kronian auroral regions, the conditions in which CMI operates and how the wave properties evolve along the raypath away from their source region. This topic is of particular interest in the frame of the search for (and the subsequent study of) CMI-driven auroral radio emissions radiated by exoplanets [Zarka et al., 2007, 2011; Griessmeier et al., this issue and references therein], brown dwarfs [e.g. Lynch et al., 2015; Enriquez et al., this issue and references therein] or cool stars [e.g. Hallinan et al., 2015 and references therein].

This article reviewed our current knowledge of general SKR properties, and pointed out a number of pending questions which summarize as :
\begin{itemize}
\item Is the emission saturated and, if so, by which process ? Is the auroral plasma homogeneous in average ?
\item What is the degree of relationship between SKR and atmospheric aurorae ? Are there radio sources conjugate to each auroral feature (such as polar spots, equatoward secondary oval etc.) and/or to upgoing electrons (as in the case of jovian S-bursts) ?
\item What is the CMI average source of free energy (electron energy, loss cone and/or shell distribution functions), the source background term and the typical size of elementary sources ?
\item How does the beaming angle vary with time, frequency and azimuth around the magnetic field vector at the source ? How does refraction close to or outside the source affect the emission angle ? 
\item What are the statistical characteristics of the polarization ellipse ? How do these vary with the source location and raypath ? 
\item What informations are ultimately carried by the wave emergence angle and polarization onto the source region and/or the crossed plasma ($e.g.$ Faraday rotation) ?
\item What is is the typical sub-corotational velocity of flux tubes carrying radio sources and producing SKR arcs ? How does SKR vary at time scales shorter than the radio period and why (source lifetime, fine structures) ? Is the planetary rotation responsible for SKR intensifications lasting for a few hours ? 
\item How and why do the SKR hemispheric periods evolve around and beyond the solstice of mid-2017 ? What is the cause of long-term (yearly) variations ? What is the driver of hemispheric rotating field-aligned currents systems ?
\end{itemize}

At the time of writing, the Cassini spacecraft just successfully ended its mission and therefore already had a number of occasion to sample SKR sources. The Cassini plasma instruments (magnetic field, particles) and spectro-imagers (UV, IR, visible) will help to analyze in situ radio measurements, while regular observations of the Hubble Space Telescope (HST) scheduling during the predicted pass within SKR sources will provide instantaneous images of the auroral context between June 2016 and September 2017. 

\begin{figure}[ht]
\centering
\includegraphics[width=1\textwidth]{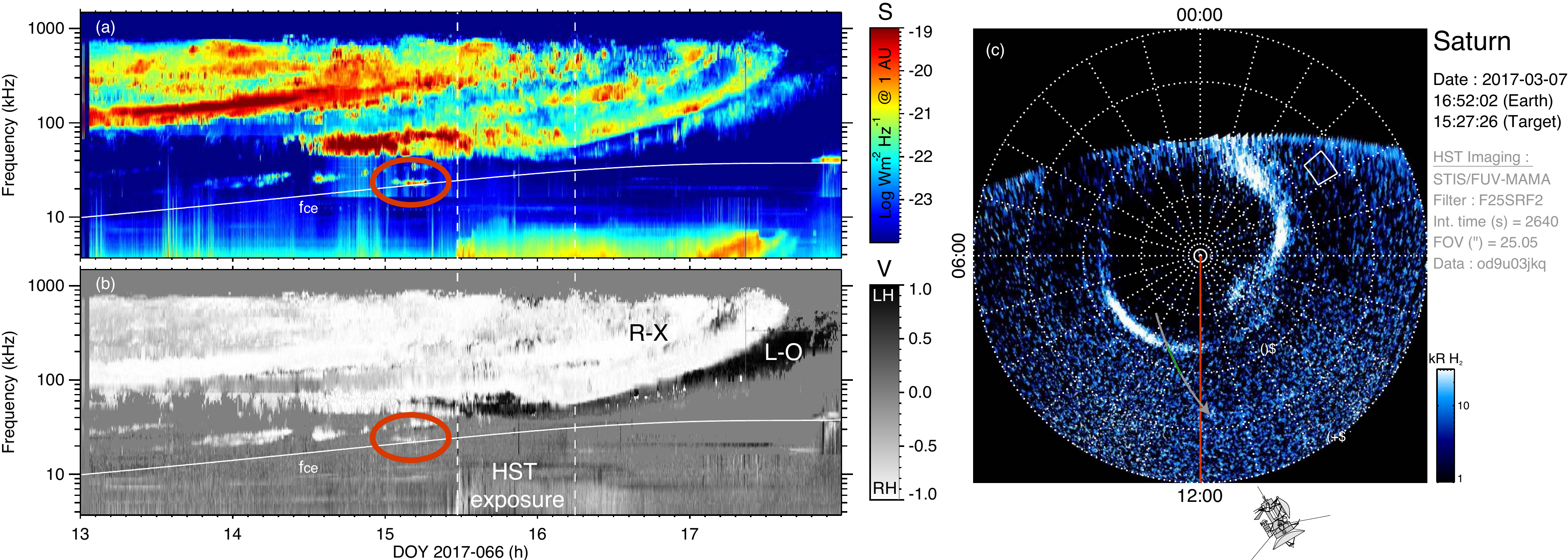}
\caption{Cassini/RPWS SKR dynamic spectrum of (a) flux density S and (b) degree of circular polarization V. The white lines draw $f_{ce}$, as derived from Cassini/MAG measurements. SKR R-X mode emission was observed at and below $f_{ce}$ within the red ellipse, indicative of a source region. (c) Polar projection of Saturn's northern aurorae imaged with HST. The gray line marks the Cassini footpath where the green portion matches the HST exposure (indicated by vertical dashed lines on left panels). The SKR source is consistent with the pass of the Cassini spacecraft along magnetic field lines connected to the main auroral oval.}
\label{fig4}
\end{figure}

Figure \ref{fig4} provides an illustrative example of a possible SKR northern source encountered on day 2017-066 15:15 at 22~kHz, 3.6 radii, 10.3~LT and 73$^\circ$ latitude.


\section*{Acknowledgements}
The author is supported by CNES and the PNP and PNST programs of CNRS/INSU and thanks the RPWS teams at both the University of Iowa and the Observatory of Meudon, the ISSI team on rotational phenomena in Saturn's magnetosphere where the work on SKR periods was discussed, and the APIS service at LESIA/Paris Astronomical Data Centre which was used to browse HST observations of GO program \#14811 concurrent with Cassini polar passes and possible SKR sources. A list of predicted source crossings, together with datasets providing Meudon SKR periods and phases discussed in this article are available at \url{http://www.lesia.obspm.fr/kronos}.


\section*{References}
\everypar={\hangindent=1truecm \hangafter=1}


Andrews,~D.\,J., A.\,J.~Coates, S.\,W.\,H.~Cowley, M.\,K.~Dougherty, L.~Lamy, G.~Provan, and P.~Zarka, Magnetospheric period oscillations at Saturn: Comparison of equatorial and high--latitude magnetic field periods with north and south Saturn kilometric radiation periods, \textsl{J. Geophys. Res.}, \textbf{115}, A12252, 2010.

Badman,~S.\,V., G.~Branduardi--Raymont, M.~Galand, S.~Hess, N.~Krupp, L.~Lamy, H.~Melin, and C.~Tao, Auroral processes at the giant planets: Energy deposition, emission mechanism, morphology and spectra, \textsl{Space Sci. Rev.}, \textbf{187}, 99--179, 2015.

Boischot,~A., Y.~Leblanc, A.~Lecacheux, B.\,M.~Pedersen, and M.\,L.~Kaiser, Arc structure in Saturn's radio dynamic spectra, \textsl{Nature}, \textbf{292}, 727--728, 1981.

Bunce,~E.\,J., S.\,W.\,H.~Cowley, D.\,L.~Talboys, M.\,K.~Dougherty, L.~Lamy, W.\,S.~Kurth, P.~Schippers, B.~Cecconi, P.~Zarka, C.\,S.~Arridge, and A.\,J.~Coates, Extraordinary field-aligned current signatures in Saturn s high-latitude magnetosphere: Analysis of Cassini data during Revolution 89, \textsl{Geophys. Res. Lett.}, \textbf{115}, A10238, 2010.

Bunce,~E.\,J., Origins of SaturnÕs Auroral Emissions and Their Relationship to Large-Scale Magnetosphere Dynamics, in \textsl{Auroral Phenomenology and Magnetospheric Processes: Earth And Other Planets}, edited by A.~Keiling, E.~Donovan, F.~Bagenal and T.~Karlsson, American Geophysical Union, Washington, D.~C., 333--374, 2012.

Carbary,~J.\,F., M.\,M.~Hedman, T.\,W.~Hill, X.~Jia, W..\,S.~Kurth, L.~Lamy, and G.~Provan, The Mysterious Periodicities of Saturn : Clues for the Rotation Rate of a Planet, in \textsl{Saturn in the 21st century}, edited by K.~Baines, M.~Flasar, N.~Krupp and T.\,S.~Stallard, Cambridge University Press, in press.

Cecconi,~B., P.~Zarka, and W.\,S.~Kurth, SKR polarization and source localization with the Cassini/RPWS/HFR instrument: First results, in \textsl {Planetary Radio Emissions VI}, edited by H.\,O.~Rucker, W.\,S.~Kurth, and G.~Mann, Austrian Academy of Sciences Press, Vienna, 37--49, 2006.

Cecconi,~B., L.~Lamy, P.~Zarka, R.~Prang\'{e}, W.\,S.~Kurth, and P.~Louarn, Goniopolarimetric study of the revolution 29 perikrone using the Cassini Radio and Plasma Wave Science instrument high-frequency radio receiver, \textsl{J. Geophys. Res.}, \textbf{114}, A03215, 2009.

Cowley,~S.~W.~H. and G.~Provan, Planetary period oscillations in SaturnÕs magnetosphere: comments on the relation between post-equinox periods determined from magnetic field and SKR emission data, \textsl{Ann. Geophys.}, \textbf{33}, 901--912, 2015.


Desch,~M.\,D., Evidence for solar wind control of Saturn radio emission, \textsl{J. Geophys. Res.}, \textbf{87}, 4549--4554, 1982.

Desch,~M.\,D., Radio emission signature of Saturn immersions in Jupiter's magnetic tail, \textsl{J. Geophys. Res.}, \textbf{88}, 6904--6910, 1983a.

Desch,~M.\,D., and H.\,O.~Rucker, The relationship between Saturn kilometric radiation and the solar wind, \textsl{J. Geophys. Res.}, \textbf{88}, 8999--9006, 1983b.

Desch,~M.\,D. and H.\,O.~Rucker, Saturn radio emission and the solar wind: Voyager 2 studies, COSPAR proceedings, \textsl{Adv. Space Res.}, \textbf{5}, 4, 333, 1985.

Farrell,~W.\,M., M.\,D.~Desch, M.\,L.~Kaiser, A.~Lecacheux, W.\,S.~Kurth, D.\,A.~Gurnett, B.~Cecconi, and P.~Zarka, A nightside source of Saturn's kilometric radiation: Evidence for an inner magnetosphere energy driver,
\textsl{Geophys. Res. Lett.}, \textbf{32}, L18107, 2005.

Fischer,~G., B.~Cecconi, L.~Lamy, S.--Y.~Ye, U.~Taubenschuss, W.~Macher, P.~Zarka, W.\,S.~Kurth, and D.\,A.~Gurnett, Elliptical polarization of Saturn kilometric radiation observed from high latitudes, \textsl{J. Geophys. Res.}, \textbf{114}, A08216, 2009.

Fischer,~G., S.--Y.~Ye, J.\,B.~Groene, A.\,P.~Ingersoll, K.\,M.~Sayanagi, J.\,D.~Menietti, W.\,S.~Kurth, and D.\,A.~Gurnett, A possible influence of the Great White Spot on Saturn kilometric radiation periodicity, \textsl{Ann. Geophys.}, \textbf{32}, 1463--1476, 2014.

Fischer,~G., D.\,A.~Gurnett, W.\,S.~Kurth, S.--Y.~Ye, and J.\,B.~Groene, Saturn kilometric radiation periodicity after equinox, \textsl{Icarus}, \textbf{254}, 72--91, 2015.

Galopeau,~P.\,H.\,M., P.~Zarka, and D.~Le Qu\'eau, Theoretical model of Saturn's kilometric radiation spectrum, \textsl{J. Geophys. Res.}, \textbf{94}, 8739--8755, 1989.

Galopeau,~P.\,H.\,M., P.~Zarka, and D.~Le Qu\'eau, Source location of Saturn's kilometric radiation: The Kelvin--Helmholtz instability hypothesis, \textsl{J. Geophys. Res.}, \textbf{100}, 26397--26410, 1995.

Galopeau,~P.\,H.\,M., and A.~Lecacheux, Variations in Saturn's radio rotation period measured at kilometer wavelengths, \textsl{J. Geophys. Res.}, \textbf{105}, 13089--13101, 2000.

Galopeau,~P.\,H.\,M. and M.\,Y.~Boudjada, An oblate beaming cone for Io-controlled Jovian decameter emission, \textsl{J. Geophys. Res.}, \textbf{121}, 3120Ð3138, 2016.

Gurnett,~D.\,A., et al. (26 co--authors), Radio and plasma wave observations at Saturn from Cassini's approach and first orbit, \textsl{Science}, \textbf{307}, 1255--1259, 2005.

Gurnett,~D.\,A., A.~Lecacheux, W.\,S.~Kurth, A.\,M.~Persoon, J.\,B.~Groene, L.~Lamy, P.~Zarka, and J.\,F.~Carbary, Discovery of a north-south asymmetry in Saturn's radio rotation period, \textsl{Geophys. Res. Lett.}, \textbf{36}, L16102, 2009.

Gurnett,~D.\,A., J.\,B.~Groene, A.\,M.~Persoon, J.\,D.~Menietti, S.-Y.~Ye, W.\,S.~Kurth, R.\,J.~MacDowall, and A.~Lecacheux, The reversal of the north and south modulation rates of the north and south components of Saturn kilometric radiation near equinox, \textsl{Geophys. Res. Lett.}, \textbf{37}, L24101, doi;1029/2010GL045796, 2010.

Gurnett,~D.\,A., J.\,B.~Groene, T.\,F.~Averkamp, W.\,S.~Kurth, S.--Y.~Ye, and G.~Fischer, A SLS4 Longitude System Based on a Tracking Filter Analysis of the Rotational Modulation of Saturn Kilometric Radiation, in \textit {Planetary Radio Emissions VII}, edited by H.\,O.~Rucker, W.\,S.~Kurth, and G.~Fischer, 51--64, 2011.

Gustin,~J., D.~Grodent, A.~Radioti, W.~Pryor, L.~Lamy and J.~Ajello, Statistical study of SaturnÕs auroral electron properties with Cassini/UVIS FUV spectral images, \textsl{Icarus}, \textbf{284}, 263--283, 2017.

Hallinan,~G. et al. (14 co--authors), Magnetospherically driven optical and radio aurorae at the end of the stellar main sequence, \textsl{Nature}, \textbf{523}, 568--571, 2015.

Hess,~S.\,L.\,G., E.~Echer, P.~Zarka, L.~Lamy, and P.\,A.~Delamere, Multi--instrument study of the Jovian radio emissions triggered by solar wind shocks and inferred magnetospheric subcorotation rates, \textsl{Planet. Space Sci.}, \textbf{99}, 136--148, 2014.

Huff,~R.\,L., W.~Calvert, J.\,D.~Craven, L.\,A.~Frank, and D.\,A.~Gurnett, Mapping of the Auroral Kilometric Radiation sources to the aurora, \textsl{J. Geophys. Res.}, \textbf{93}, 11445, 1988.

Hunt,~G.~J. et al. (8 co-authors), Field-aligned currents in SaturnÕs southern nightside magnetosphere: Subcorotation and planetary period oscillation components, \textsl{J. Geophys. Res.}, \textbf{119}, 9847--9899, 2014.

Jia,~X., M.~G.~Kivelson and T.~I.~Gombosi, Driving SaturnÕs magnetospheric periodicities from the upper atmosphere/ionosphere, \textsl{J. Geophys. Res.}, \textbf{117}, A04215, 2012.

Jackman,~C.\,M., L.~Lamy, M.\,P.~Freeman, P.~Zarka, B.~Cecconi, W.\,S.~Kurth, S.\,W.\,H.~Cowley, and M.\,K.~Dougherty, On the character and distribution of lower-frequency radio emissions at Saturn and their relationship to substorm-like events, \textsl{J. Geophys. Res.}, \textbf{114}, A08211, doi:10.1029/2008JA013997, 2009.

Jackman,~C.\,M., C.\,S.~Arridge, J.\,A.~Slavin, S.\,E.~Milan, L.~Lamy, M.\,K.~Dougherty, and A.\,J.~Coates, In situ observations of the effect of a solar wind compression on Saturn's magnetotail, \textsl{J. Geophys. Res.}, \textbf{115}, A10240, doi:10.1029/2010JA015312, 2010.

Jackman,~C.\,M., Saturn radio emissions and their relation to magnetospheric dynamics, in \textsl{Planetary Radio Emissions VII}, edited by H.\,O.~Rucker, W.\,S.~Kurth, P.~Louarn, and G.~Fischer, Austrian Academy of Sciences Press, Vienna, 1--12, 2011.

Kaiser,~M.\,L., M.\,D.~Desch, J.\,W.~Warwick, and J.\,B.~Pearce, Voyager detection of nonthermal radio emission from Saturn, \textsl{Science}, \textbf{209}, 1238--1240, 1980.

Kaiser,~M.\,L., and M.\,D.~Desch, Saturnian kilometric radiation - Source locations, \textsl{J. Geophys. Res.}, \textbf{87}, 4555--4559, 1982.

Kaiser,~M.\,L., M.\,D.~Desch, and A.~Lecacheux, Saturnian kilometric radiation: Statistical properties and beam geometry, \textsl{Nature}, \textbf{292}, 731--733, 1981.

Kaiser,~M.\,L., M.\,D.~Desch, W.\,S.~Kurth, A.~Lecacheux, F.~Genova, B.\,M.~Pedersen, and D.\,R.~Evans, Saturn as a radio source, in \textsl{Saturn}, edited by T.~Gehrels, and M.\,S.~Matthews, University of
Arizona Press, Tucson, USA, 378--415, 1984.

Kimura,~T., L.~Lamy, C.~Tao, S.\,V.~Badman, S.~Kasahara, B.~Cecconi, P.~Zarka, A.~Morioka, Y.~Miyoshi, D.~Maruno, Y.~Kasaba, and M.~Fujimoto, Long-term modulations of SaturnÕs auroral radio emissions by the solar wind and seasonal variations controlled by the solar ultraviolet flux, \textsl{J. Geophys. Res.}, \textbf{118}, 7019--7035, 2013.

Kopf,~A., D.\,G.~Gurnett, A Statistical Study of Kilometric Radiation Fine Structure Striations Observed at Jupiter and Saturn, \textsl{Geophysical Research Abstracts, EGU General Assembly, Austria}, \textbf{13}, EGU2011-8726, 2011.

Kurth,~W.\,S., G.\,B.~Hospodarsky, D.\,A.~Gurnett, B.~Cecconi, P.~Louarn, A.~Lecacheux, P.~Zarka, H.\,O.~Rucker, M.~Boudjada, and M.\,L.~Kaiser, High spectral and temporal resolution observations of Saturn kilometric radiation, \textsl{Geophys. Res. Lett.}, \textbf{32}, L20S07, 2005.

Kurth,~W.\,S., E.\,J.~Bunce, J.\,T.~Clarke, F.\,J.~Crary, D.\,C.~Grodent, A.\,P.~Ingersoll, U.\,A.~Dyudina, L.~Lamy, D.\,G.~Mitchell, A.\,M.~Persoon, W.\,R.~Pryor, J.~Saur and T.~Stallard, Auroral processes, in \textsl{Saturn from Cassini--Huygens}, edited by M.\,K.~Dougherty, L.\,W.~Esposito, and S.\,M.~Krimigis, Springer, Netherlands, 333--374, 2009.

Kurth,~W.\,S., et al. (20 co--authors), A close encounter with a Saturn kilometric radiation source region, \textsl{Planetary Radio Emissions VII}, edited by H.\,O.~Rucker, W.\,S.~Kurth, P.~Louarn, and G.~Fischer, Austrian Acad. Sci. Press, Vienna, 75--85, 2011.

Kurth,~W.\,S., G.\,B.~Hospodarsky, D.\,A.~Gurnett, L.~Lamy, M.\,K.~Dougherty, J.~Nichols, E.\,J.~Bunce, W.~Pryor, K.~Baines, T.~Stallard, H.~Melin, and F.\,J.~Crary, Saturn kilometric radiation intensities during the Saturn auroral campaign of 2013, \textsl{Icarus}, \textbf{263}, 2--9, 2016.

Lamy,~L., P.~Zarka, B.~Cecconi, R.~Prange, W.\,S.~Kurth, and D.\,A.~Gurnett, Saturn kilometric radiation: Average and statistical properties, \textsl{J. Geophys. Res.}, \textbf{113}, A07201, 2008a.

Lamy,~L., P.~Zarka, B.~Cecconi, S.~Hess and R.~Prang\'e, Modeling of Saturn kilometric radiation arcs and equatorial shadow zone, \textsl{J. Geophys. Res.}, \textbf{113}, A10213, 2008b.

Lamy,~L., Study of radio auroral emissions from Saturn, modeling and UV aurorae, PhD Thesis, Universit\'e Pierre et Marie Curie, Observatoire de Paris, Meudon, 9 Sept. 2008c.

Lamy,~L., B.~Cecconi, R.~Prange, P.~Zarka, J.\,D.~Nichols, and J.\,T.~Clarke, An auroral oval at the footprint of Saturn's kilometric radio sources, colocated with the UV aurorae, \textsl{J. Geophys. Res.}, \textbf{114}, A10212, 2009.

Lamy,~L., P.~Schippers, P.~Zarka, B.~Cecconi, C.\,S.~Arridge, M.\,K.~Dougherty, P.~Louarn, N.~André, W.\,S.~Kurth, R.\,L.~Mutel, D.\,A.~Gurnett, and A.\,J.~Coates, Properties of Saturn kilometric radiation measured within its source region, \textsl{Geophys. Res. Lett.}, \textbf{37}, L12104, 2010.

Lamy,~L., B.~Cecconi, P.~Zarka, P.~Canu, P.~Schippers, W.\,S.~Kurth, R.\,L.~Mutel, D.\,A.~Gurnett, D.~Menietti, and P.~Louran, Emission and propagation of Saturn kilometric radiation: Magnetoionic modes, beaming pattern, and polarization state, \textsl{J. Geophys. Res.}, \textbf{116}, A04212, 2011.

Lamy,~L., Variability of southern and northern SKR periodicities, in \textsl{Planetary Radio Emissions VII}, edited by H.\,O.~Rucker, W.\,S.~Kurth, P.~Louarn, and G.~Fischer, Austrian Academy of Sciences Press, Vienna, 39--50, 2011.

Lamy,~L., R.~Prang\'e, W.~Pryor, J.~Gustin, S.\,V.~Badman, H.~Melin, T.~Stallard, D.\,G.~Mitchell, and P.\,C.~Brandt, Multispectral simultaneous diagnosis of Saturn's aurorae throughout a planetary rotation, \textsl{J. Geophys. Res.}, \textbf{118}, 4817--4843, 2013.

Lecacheux,~A., and F.~Genova, Source location of Saturn kilometric radio emission, \textsl{J. Geophys. Res.}, \textbf{88}, 8993--8998, 1983.

Lynch,~C., R.\,L.~Mutel and M.~Gudel, Wideband dynamic radio spectra of two ultra-cool dwarfs, \textsl{Astrophys. J.}, 802--106, 2015.

Melrose,~D.\,B., and G.\,A.~Dulk, On the elliptical polarization of Jupiter's decametric radio emission, \textsl{Astron. Astrophys.}, \textbf{249}, 250--257, 1991.

Menietti,~J.\,D., R.\,L.~Mutel, P.~Schippers, S.--Y.~Ye, O.~Santolik, W.\,S.~Kurth, D.\,A.~Gurnett, L.~Lamy, and B.~Cecconi, Saturn kilometric radiation near a source center on day 73, 2008, in \textsl{Planetary Radio Emissions VII}, edited by H.\,O.~Rucker, W.\,S.~Kurth, P.~Louarn, and G.~Fischer, Austrian Academy of Sciences Press, Vienna, 87--95, 2011.

Menietti,~J.\,D., R.\,L.~Mutel, P.~Schippers, S.--Y.~Ye, D.\,A.~Gurnett, and L.~Lamy, Analysis of Saturn kilometric radiation near a source center, \textsl{J. Geophys. Res.}, \textbf{116}, A12222, 2011.

Mutel,~R.\,L., I.\,W.~Christopher, and J.\,S.~Pickett, Cluster multispacecraft determination of AKR angular beaming, \textsl{Geophys. Res. Lett.}, \textbf{37}, L07104, 2008.

Mutel,~R.\,L., J.\,D.~Menietti, D.\,A.~Gurnett, W.~Kurth, P.~Schippers, C.~Lynch, L.~Lamy, C.~Arridge, and B.~Cecconi, CMI growth rates for Saturnian kilometric radiation, \textsl{Geophys. Res. Lett.}, \textbf{37}, L19105, 2010.

Ortega--Molina,~A., and A.~Lecacheux, Polarization response of the Voyager-PRA experiment at low frequencies, \textsl{Astron. Astrophys.}, \textbf{229}, 558--568, 1990.

Provan,~G., L.~Lamy, S.\,W.\,H.~Cowley, and M.\,K.~Dougherty, Planetary period oscillations in Saturn's magnetosphere: Comparison of magnetic oscillations and SKR modulations in the postequinox interval, \textsl{J. Geophys. Res.}, \textbf{119}, 7380--7401, 2014.

Provan,~G., C.~Tao, S.~W.~H.~Cowley, M.~K.~Dougherty, and A.~J.~Coates, Planetary period oscillations in SaturnÕs magnetosphere: Examining the relationship between abrupt changes in behavior and solar wind-induced magnetospheric compressions and expansions, \textsl{J. Geophys. Res.}, \textbf{120}, 9524--9544, 2015.

Provan,~G., S.\,W.\,H.~Cowley, L.~Lamy, E.\,J.~Bunce, G.\,J.~Hunt, P.~Zarka and M.\,K.~Dougherty, Planetary period oscillations in SaturnÕs magneto- sphere: Coalescence and reversal of northern and southern periods in late northern spring, \textsl{J. Geophys. Res.}, \textbf{121}, 2016.

Rucker,~H.\,O., M.~Panchenko, K.\,C.~Hansen, U.~Taubenschuss, M.\,Y.~Boudjada, W.\,S.~Kurth, M.\,K.~Dougherty, J.\,T.~Steinberg, P.~Zarka, P.\,H.\,M.~Galopeau, D.\,J.~McComas, and C.\,H.~Barrow, Saturn
kilometric radiation as a monitor for the solar wind? \textsl{Adv. Space Res.}, \textbf{42}, 40--47, 2008.

Schippers,~P., C.\,S.~Arridge, J.\,D.~Menietti, D.\,A.~Gurnett, L.~Lamy, B.~Cecconi, D.\,G.~Mitchell, N.~Andre, W.\,S.~Kurth, S.~Grimald, M.\,K.~Dougherty, A.\,J.~Coates, and D.\,T.~Young, Auroral electron distributions within and close to the Saturn kilometric radiation source region, \textsl{J. Geophys. Res.}, \textbf{166}, A05203, 2011.

Southwood,~D.\,J., and S.\,W.\,H.~Cowley, The origin of SaturnÕs magnetic periodicities: Northern and southern current systems, \textsl{J. Geophys. Res.}, \textbf{119}, 1563--1571, 2014.

Stallard,~T.\,S., S.\,V.~Badman, U.\,A.~Dyudina, D.\,C.~Grodent and L.~Lamy, Saturn's aurora, in \textsl{Saturn in the 21st century}, edited by K.~Baines, M.~Flasar, N.~Krupp and T.\,S.~Stallard, Cambridge University Press, in press.

Thieman,~J.\,R., and M.\,L.~Goldstein, Arcs in Saturn's radio spectra, \textsl{Nature}, \textbf{292}, 728--731, 1981.

Trauger,~J.\,T., et al. (16 co--authors), Saturn's hydrogen aurora: Wide field and planetary camera 2 imaging from the Hubble Space Telescope, \textsl{J. Geophys. Res.}, \textbf{103}, 20237--20244, 1998.

Treumann,~R.\,A., The electron--cyclotron maser for astrophysical application, \textsl{Astron. Astrophys. Rev.}, \textbf{13}, 229--315, 2006.

Warwick,~J.\,W., J.\,B.~Pearce, D.\,R.~Evans, T.\,D.~Carr, J.\,J.~Schauble, J.\,K.~Alexander, M.\,L.~Kaiser, M.\,D.~Desch, B.\,M.~Pedersen, A.~Lecacheux, G.~Daigne, A.~Boischot, and C.\,H.~Barrow,
Planetary Radio Astronomy observations from Voyager~1 near Saturn, \textsl{Science}, \textbf{212}, 239--243, 1981.

Warwick,~J.\,W., D.\,R.~Evans, J.\,H.~Romig, J.\,K.~Alexander, M.\,D.~Desch, M.\,L.~Kaiser, M.~Aubier, Y.~Leblanc, A.~Lecacheux, and B.\,M.~Pedersen, Planetary Radio astronomy Observations from Voyager~2 near Saturn, \textsl{Science}, \textbf{215}, 582--587, 1982.

Wu,~C.\,S., and L.\,C.~Lee, A theory of terrestrial kilometric radiation, \textsl{Astrophys. J.}, \textbf{230}, 621--626, 1979.

Ye,~S.--Y., G.~Fischer, W.\,S.~Kurth, J.\,D.~Menietti, and D.\,A.~Gurnett, Rotational modulation of Saturn's radio emissions after equinox, \textsl{J. Geophys. Res.}, \textbf{121}, 11714--11728, 2016.

Zarka,~P., Auroral radio emissions at the outer planets: Observations and theories, \textsl{J. Geophys. Res.}, \textbf{103}, 20159--20194, 1998.

Zarka,~P., Plasma interactions of exoplanets with their parent star and associated radio emissions, \textsl{Planet. Space Sci.}, \textbf{55}, 598--617, 2007.

Zarka,~P., The search for exoplanetary radio emissions, in \textsl{Planetary Radio Emissions VII}, edited by H.\,O.~Rucker, W.\,S.~Kurth, P.~Louarn, and G.~Fischer, Austrian Academy of Sciences Press, Vienna, 287--301, 2011.

\end{document}